\documentclass[aps,prb,reprint,showpacs]{revtex4-1}
\usepackage{amsmath,amssymb}
\usepackage[dvips]{graphicx}

\usepackage{bm}

\usepackage[usenames]{color} 

\begin{document}

\title{In-plane magnetic field dependence of cyclotron relaxation time\\
in a Si two-dimensional electron system}

\author{Tasuku Chiba,$^1$ Ryuichi Masutomi,$^1$
Kentarou Sawano,$^2$ Yasuhiro Shiraki,$^2$ and Tohru Okamoto$^1$}
\affiliation{$^1$Department of Physics, University of Tokyo, 7-3-1, Hongo, Bunkyo-ku, Tokyo 113-0033, Japan\\
$^2$Research Center for Silicon Nano-Science, Advanced Research Laboratories, Tokyo City University,
8-15-1 Todoroki, Setagaya-ku, Tokyo 158-0082, Japan}

\date{Received 15 February 2012; published 13 July 2012}

\begin{abstract}

Cyclotron resonance of two-dimensional electrons is studied
for a high-mobility Si/SiGe quantum well
in the presence of an in-plane magnetic field,
which induces spin polarization.
The relaxation time $\tau_{\rm CR}$ shows a negative in-plane magnetic field dependence,
which is similar to that of the transport scattering time $\tau_t$ obtained
from dc resistivity.
The resonance magnetic field shows an unexpected negative shift with increasing in-plane magnetic field.
\end{abstract}

\maketitle

Low-density and strongly correlated two-dimensional (2D) systems have attracted much attention.
\cite{Abrahams2001,Sarma2005,Spivak2010}
Metallic temperature dependence of resistivity $\rho(T)$ has been observed in 2D systems
where a Wigner-Seitz radius $r_s$ is much larger than unity.
Extensive experimental and theoretical studies have been carried out,
as this metallic behavior contradicts the scaling theory of localization in two dimensions.
\cite{Abrahams1979}
However, the origin of the metallic behavior remains unclear and controversial.
Another intriguing feature of low density 2D systems is
a dramatic response to a magnetic field parallel to the 2D plane.
Strong positive magnetoresistance has been reported for 
various 2D carrier systems such as Si-metal oxide semiconductor field effect transistors (MOSFETs),
\cite{Simmonian1997,Okamoto1999}
$n$-Si quantum wells (QWs),
\cite{Okamoto2000,Okamoto2004,Lai2005}
$p$-GaAs/AlGaAs,
\cite{Yoon2000,Tutuc2001}
$n$-GaAs/AlGaAs heterojunctions,
\cite{Tutuc2002,Zhu2003}
$n$-AlAs,
\cite{Poortere2002,Gunawan2007}
$p$-GaAs,
\cite{Gao2006}
and
$p$-SiGe QWs.
\cite{Drichko2009}
It is related to the spin polarization $P$
since an in-plane magnetic field $B_\parallel$ does not couple to the 2D motion of carriers.
\cite{Okamoto1999}
The positive dependence of $\rho$ on $P$ is also a subject of discussion.

Recently, Masutomi \textit{et al.} have performed cyclotron resonance (CR) measurements
on high-mobility Si 2D electron systems (2DESs).
\cite{Masutomi2011}
The relaxation time $\tau_{\rm CR}$, obtained from the linewidth,
was found to be comparable to the transport scattering time $\tau_t$.
It increases with decreasing temperature in a fashion similar to $\tau_t$.
The results indicate that the scattering time has the metallic $T$ dependence
over a wide frequency range.
In this work, we study $\tau_{\rm CR}$ in the presence of an in-plane magnetic field.
It decreases as $B_\parallel$ increases.
The $B_\parallel$ dependence is also similar to that of $\tau_t$,
which corresponds to the positive magnetoresistance. 

The sample was fabricated from the same wafer as the one studied in Ref.~\onlinecite{Masutomi2011}.
It is a Si/SiGe heterostructure with a 20-nm-thick strained Si QW
sandwiched between relaxed $\mathrm{Si}_{0.8}\mathrm{Ge}_{0.2}$ layers.
\cite{Yutani1996}
The electrons are provided by a Sb-$\delta$-doped layer 20~nm above the channel.
The 2D electron concentration $N_s$ was adjusted to 1.13 $\times 10^{15}~\mathrm{m^{-2}}$
at 20~K with bias voltage $V_{\rm BG} = -5.5$~V
of a $p$-type Si(001) substrate 2.1~$\mu$m below the channel.
A two-axis vector magnet system was used to apply independently $B_\parallel$ and
the perpendicular magnetic field $B_\perp$ for CR measurements.
Instead of using a bolometer,\cite{Masutomi2011}
we observe electron heating in the 2DES
under excitation at 100~GHz.
A Hall bar sample, whose channel width is 200~$\mu$m,
was mounted inside an oversized waveguide with an 8-mm bore
inserted into a liquid-helium cryostat.

Figure 1 shows $B_\parallel$ dependence of $\rho$ at $B_\perp=0$ for different $T$
without millmeter-wave radiation.
\begin{figure}[b]
\includegraphics[width=.9\linewidth]{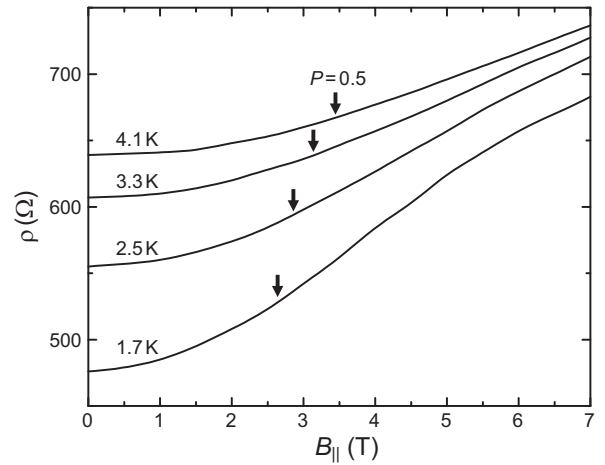}
\caption{
Resistivity $\rho$ as a function of $B_\parallel$ at $B_\perp=0$
for different temperatures.
Calculated values of $B_\parallel$ for $P=0.5$ are indicated by arrows.
}
\end{figure}
The in-plane magnetic field is oriented along the Hall bar direction.
The critical magnetic field for the full spin polarization ($P=1$) at $T=0$
is estimated to be 5.0 T at this density.
\cite{Okamoto2004}
Arrows indicate calculated values of $B_\parallel$ for $P=0.5$ at each temperature.\cite{OurCal}
While $P$ decreases with increasing $T$ for a fixed $B_\parallel$,
the reduction of $P$ is not significant in this temperature range.
In contrast to the case of Si-MOSFETs
\cite{Simmonian1997,Okamoto1999}
and $p$-GaAs/AlGaAs heterojunctions,
\cite{Yoon2000,Tutuc2001}
high-mobility Si 2DESs exhibit apparent metallic behavior 
even in the spin-polarized regime.
\cite{Okamoto2000,Okamoto2004,Lai2005}
In Ref.~\onlinecite{Okamoto2004},
Okamoto \textit{et al.} proposed a schematic phase diagram and
pointed out the importance of low disorder and the valley degree of freedom.
The essential role of the valley degree of freedom for the metallic behavior
is also reported for an AlAs 2DES.
\cite{Gunawan2007}

Figure 2(a) shows $B_\perp$ dependence of the longitudinal resistivity $\rho_{xx}$
at $B_\parallel=3.0$~T.
The lattice or bath temperature, $T_{\rm L}$, was kept at 1.7~K
and the magnetoresistance curves were obtained with and without electromagnetic-wave excitation.
We also measured the magnetoresistance at higher temperatures
and confirmed that $T$ dependence of $\rho_{xx}$ remains metallic in the measurement region.
The radiation-induced increase in $\rho_{xx}$ can be attributed to electron heating.
The electron temperature $T_e$ is evaluated from the data obtained without radiation
for $T_e=T_{\rm L} > 1.7~{\rm K}$.
\begin{figure}[t]
\includegraphics[width=1.0\linewidth]{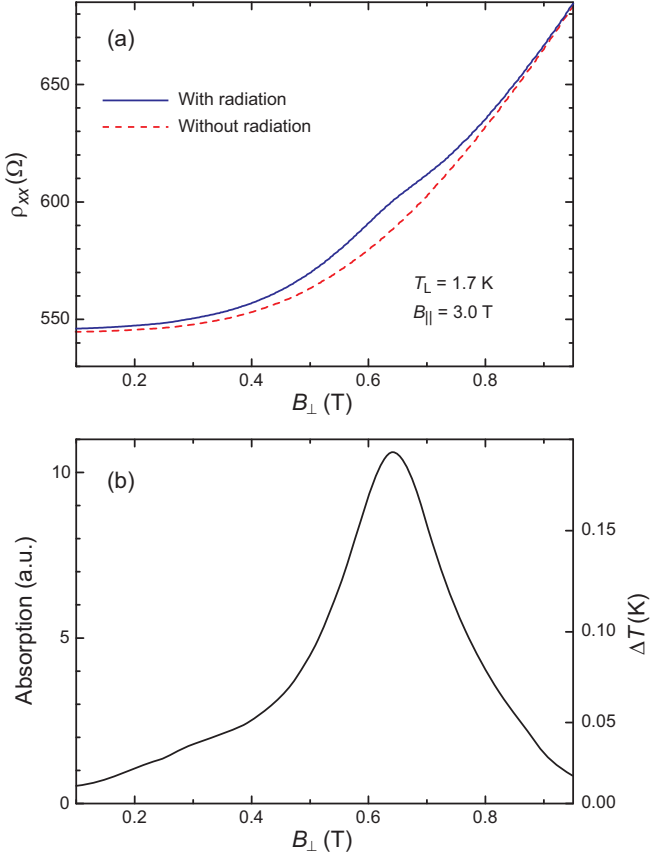}
\caption{(color online).
(a) $B_\perp$ dependence of the longitudinal resistivity $\rho_{xx}$,
with (solid curve) and without (dashed curve) electromagnetic-wave excitation.
The in-plane magnetic field and the lattice temperature are fixed
at $B_\parallel =3.0$~T and $T_{\rm L}=1.7$~K, respectively.
(b) CR trace deduced from the electron temperature $T_e$.
The energy absorption per unit area is assumed to be proportional to
$T_e^5-T_{\rm L}^5$.
The corresponding temperature difference, $\Delta T=T_e-T_{\rm L}$,
is indicated on the right axis.
}
\end{figure}
Electron cooling to the lattice is expected to occur
via electron-phonon coupling.
Heat transfer rate can be obtained experimentally from
the dc current-voltage characteristics.
\cite{Prus2002,Gao2005}
Using a sample fabricated from the same wafer,
Toyama \textit{et al.} found a $T^5$ power law in the range from 0.6 to 8~K.
\cite{Toyama2006}
The relaxation time is calculated to be 2~ns at 1.7~K,
which is much longer than $\tau_t$, $\tau_{\rm CR}$, the period of the electromagnetic wave ($=10$~ps),
and the dephasing time ($\sim \hbar/k_B T=4$~ps).
The weakness of the electron-phonon coupling ensures well-defined steady-state temperature $T_e$ of the electron system.
In Fig.~2(b), the CR absorption is shown assuming that 
it is proportional to $T_e^5-T_{\rm L}^5$.
Since the radiation power was kept low so that
temperature difference $\Delta T=T_e-T_{\rm L}$ is much smaller than $T_{\rm L}$,
the CR absorption is almost proportional to $\Delta T$.

Figure 3(a) shows $B_\parallel$ dependence of $\tau_t$ and $\tau_{\rm CR}$ at $T=1.7$~K.
The transport scattering time $\tau_t=m^\ast /e^2 N_s \rho$ is determined from the data shown in Fig.~1
with the effective mass $m^\ast=0.19m_e$.
From the half width at half maximum $\Delta B$ of the CR absorption line,
the relaxation time is obtained as $\tau_{\rm CR}=B_{\rm CR}/(\omega \Delta B)$.
Here $B_{\rm CR}$ is the resonance magnetic field
and $\omega$ is the microwave frequency ($\omega/2\pi=100$~GHz).
\begin{figure}[t]
\includegraphics[width=.9\linewidth]{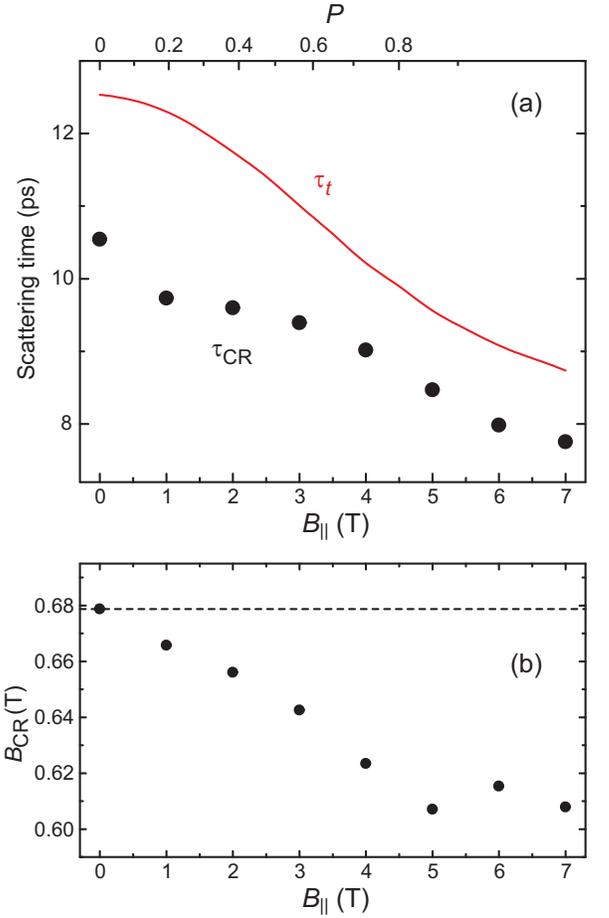}
\caption{(color online).
(a) Scattering times $\tau_t$ and $\tau_{\rm CR}$ at $T=1.7$~K as a function of $B_\parallel$.
The corresponding spin polarization is indicated on the upper axis.
(b) The resonance magnetic field $B_{\rm CR}$ versus $B_\parallel$.
The dotted line represents $B_{\rm CR}=0.19m_e \omega/e$.
}
\end{figure}
Although the data of Fig.~3(a) were obtained for a constant $T_{\rm L}$,
the difference in $T_e$ for different $B_\parallel$ is small
since $\Delta T$ was kept low and about 0.2~K at the peak for all $B_\parallel$.
The obtained $\tau_{\rm CR}$ exhibits a negative dependence on $B_\parallel$.
It is similar to that of $\tau_t$, corresponding to the positive magnetoresistance.
This suggests that the scattering time has a negative dependence on the spin polarization
over a very wide frequency range from dc to 100~GHz.
We believe that the present results, together with those of Ref.~\onlinecite{Masutomi2011},
will provide a strong constraint on theoretical models.

In Fig.~3(b), $B_{\rm CR}$ is plotted as a function of $B_\parallel$.
Unexpectedly, $B_{\rm CR}$ deviates from $0.19m_e \omega /e$ and decreases as $B_\parallel$ increases. 
Electron-spin-resonance measurements demonstrate that
the spin-orbit interactions are very small in the present system.
\cite{Matsunami2006}
An in-plane magnetic field can modify the wave function in the confinement direction
and cause a distortion of the 2D Fermi lines.
\cite{Smrcka1994}
However, this effect increases $B_{\rm CR}$.
The enhancement of the cyclotron mass
is estimated to be only about 1~\% for 7~T in the present 2DES.
\cite{Mass}

In summary, we have performed the cyclotron resonance measurements
on a high-mobility Si 2DES
in the presence of an in-plane magnetic field $B_\parallel$.
The relaxation time $\tau_{\rm CR}$, obtained from the linewidth,
was found to have a negative $B_\parallel$ dependence,
which is similar to that of the transport scattering time $\tau_t$.
The resonance peak shifts unexpectedly toward lower $B_\perp$ as $B_\parallel$ increases.

{\it Acknowledgment.}
This work was partly supported by a Grant-in-Aid for Scientific Research (A) (Grant No. 21244047)
from the Ministry of Education, Culture, Sports, Science, and Technology, Japan.


\begin{thebibliography}{99}
\bibitem{Mass}
Due to a one-side doping and applied gate voltage,
the confining potential is regarded as a triangular well
rather than a square well.
The average distance of electrons from the interface
is calculated to be 4~nm which is smaller than
the magnetic length $l_B =(\hbar/eB)^{1/2}$
(10~nm at 7~T).
\end{thebibliography}
\end{document}